\documentclass[12pt,a4paper]{article}
\usepackage{graphicx}
\usepackage{amsmath}
\usepackage{amssymb}
\usepackage{epsfig}
\usepackage{times}
\usepackage{latexsym}
\usepackage{array}
\begin{document}
%\begin{titlepage}

\title{On the first Solvay Congress in 1911}
\author{Norbert Straumann \\
Institute for Theoretical Physics University of Z\"urich,\\
Winterthurerstrasse 190, CH--8057 Z\"urich, Switzerland}
%\date{}
\maketitle
\begin{abstract}
Late in October 1911, eighteen leading scientists from all over Europe met to the first of a famous sequence of  Solvay conferences in Brussels. This historical meeting was mainly  devoted to  " The Theory of Radiation and the Quanta", at a time when the foundations of physics were totally shaken. Although ``nothing positive came out'' (Einstein), it is interesting to see the diverging attitudes of Europe's most famous scientists in the middle of the quantum revolution.  After a few general remarks about the conference, I shall focus on some of the most interesting contributions and discussions. Einstein, at 32 the youngest, was clearly most aware of the profound nature of the crises. He gave the final talk entitled ``The Present State of the Problem of Specific Heats'', but he put his theme into the larger context of the quantum problem, and caused a barrage of challenges, in particular from Lorentz, Planck, Poincar\'{e}, and others.
\end{abstract}

\section{Introduction}

The Belgian chemist Ernest Solvay (1839-1922) became a very rich man with his invention of industrial  soda production. Because he was convinced that his odd ideas on natural science, physics in particular, were important, he was eager to discuss these with some of Europe's top physicists. Walther Nernst made clever use of this interest, and suggested that Solvay may fund an elite gathering at which leading scientists would listen to his ideas on gravitation, Brownian motion, radioactivity, etc. 

\begin{figure}
\begin{center}
\includegraphics[height=0.4\textheight]{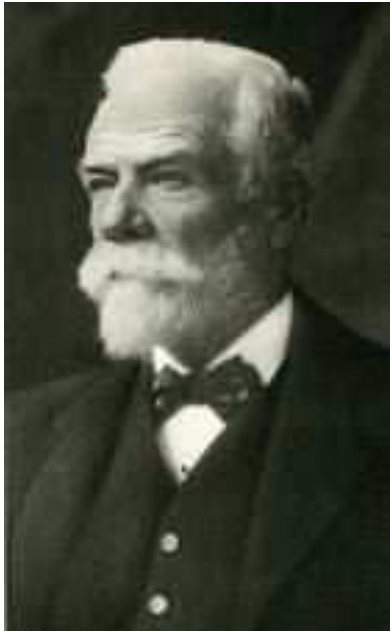}
\caption{Ernest Solvay (1839-1922).} \protect\label{ES:Fig-1}
\end{center}
\end{figure}

Late in October eighteen of Europe's most famous scientists gathered in the Grand Hotel Metropole in Brussels to discuss ``The Theory of Radiation and the Quanta'', that seemed to fundamentally overshadow classical physics. The fol\-lowing list of invited people\footnote{Lord Rayleigh and J. D. van der Waals were not present at the meetings. A letter by Rayleigh is included in the proceedings of the conference.} is indeed impressive:
\vspace{1cm}

 \hspace{3 cm}     H. A. Lorentz (Leiden), as Chairman.

 \hspace{2 cm}  \textit{From Germany}:

 \hspace{4 cm}	   W. Nernst (Berlin)
 	
 \hspace{4 cm}	   M. Planck (Berlin)
    
  \hspace{4 cm}  H. Rubens (Berlin)
     
 \hspace{4 cm}	   A. Sommerfeld (M\"{u}nchen)
     
  \hspace{4 cm}  W. Wien (W\"{u}rtzburg)
     
  \hspace{4 cm}  E. Warburg (Charlottenburg).
     
  \hspace{2 cm}  \textit{From England}:

  \hspace{4 cm}	Lord Rayleigh (London)
                                            
  \hspace{4 cm}	J. H. Jeans (Cambridge)
	
  \hspace{4 cm}   E. Rutherford (Manchester) 
  
  \hspace{2 cm}  \textit{From France}:
  
  \hspace{4 cm}  M. Brillouin (Paris)
  
  \hspace{4 cm}  Madame Curie (Paris)
  
  \hspace{4 cm}  P. Langevin (Paris)
  
  \hspace{4 cm}  J. Perrin (Paris)
  
  \hspace{4 cm} H. Poincar\'{e} (Paris)
  
  \hspace{2 cm}  \textit{From Austria}: 
  
  \hspace{4 cm}  A. Einstein (Prag)
  
  \hspace{4 cm}  F. Hasen\"{o}hrl (Vienna)
  
  \hspace{2 cm}  \textit{From Holland}: 
    
  \hspace{4 cm}  H. Kamerlingh Onnes (Leiden)
    
  \hspace{4 cm}   J. D. van der Waals (Amsterdam)
   
  \hspace{2 cm}  \textit{From Denmark}:
   
  \hspace{4 cm}  M. Knudsen (Copenhagen)
  
 \vspace{1cm} 
 
 \begin{figure}[h]
\begin{center}
\includegraphics[height=0.4\textheight]{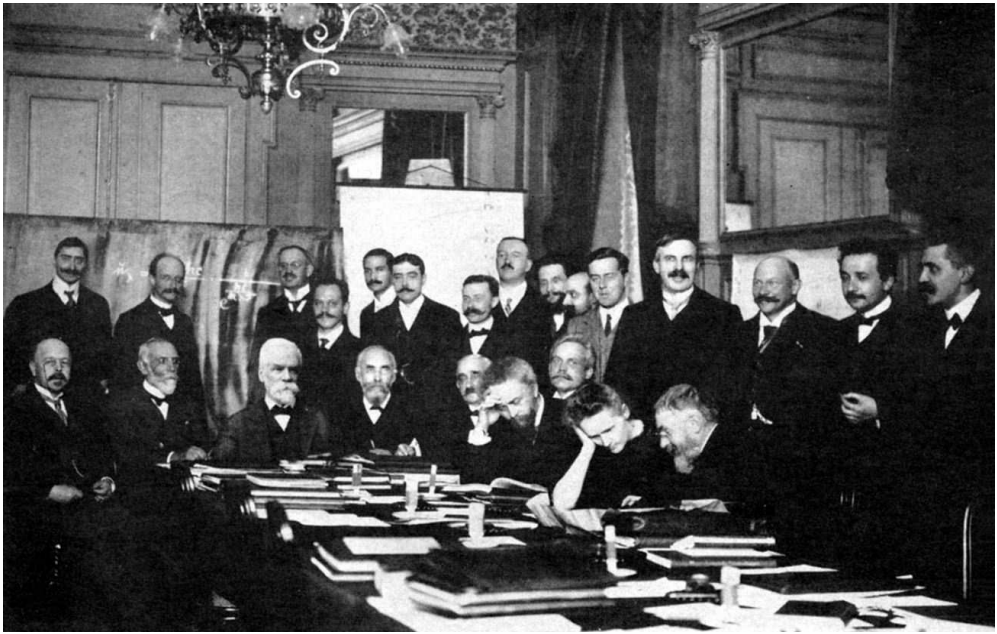}
\caption{The portrait of participants to the first Solvay Conference in 1911.}
\protect\label{SC:Fig-2}
\end{center}
\end{figure}

{\footnotesize\textit{Notes} on Fig. 2. Left to right seated: Walter Nernst; Marcel-Louis Brillouin; E. Solvay; Hendrik Lorentz; Emil Warburg; Jean-Baptiste Perrin; Wilhelm Wien; Marie Curie; Henri Poincar\'{e}. Left to right standing: Robert Goldschmidt; Max Planck; Heinrich Rubens; Arnold Sommerfeld; Frederick Lindemann; Maurice de Broglie; Martin Knudsen; Friedrich Hasen\"{o}hrl; G. Hostelet; E. Herzen; Sir James Jeans; Ernest Rutherford; Heike Kamerlingh-Onnes; Albert Einsten; Paul Langevin. Further remarks: M. de Broglie (the elderly brother of Louis de Broglie), F. Lindemann and R.B. Goldschmidt were appointed as secretaries;  G. Hostelet and E. Herzen were co-workers of E. Solvay. Solvay was not present at the time the photo was taken; his photo was pasted onto this one for the official release (resulting in a rather big head).}

During the preparation phase of the congress, with W. Nernst as its organizer, several participants were asked to write detailed reports, that were then sent in advance to the invited members. These formed the basis of the discussions, and also the main part of the proceedings \cite{Euck}. The authors that contributed are: Lorentz, Jeans, Warburg, Rubens, Planck, Knudsen, Perrin, Nernst, Kamerlingh Onnes, Sommerfeld, Langevin, and Einstein. The discussions on these reports, that are often of great historical interest, are also fully included in the proceedings. For today's physics teachers, I recommend especially Perrin's  beautiful lecture, which is an exhaustive review of the experimental evidence in favor of the existence of atoms.\footnote{In 1913 Perrin published a more extended version in his his classic book ``Les Atomes'' \cite{perr}.} Some of the reports are in my opinion not of great historical interest. For example, in his lengthy contribution Sommerfeld introduced a version of the quantum hypothesis, which he considered to be compatible with \textit{classical} electrodynamics.\footnote{Title of Sommerfeld's report:  ``Die Bedeutung des Wirkungsquantums  f\"{u}r unperiodische Molekularprozesse in der Physik'',  (\cite{Euck},  pp. 252-317). [The significance of the of action quantum for non-periodic processes in physics.] } He postulated that in ``every purely molecular process'' the quantity of action
\[\int_{0}^{\tau}\, L\,dt =\frac{h}{2\pi}\]
is exchanged, where $\tau$ is the duration of the process and $L$ the Lagrangian. Then, he applied this hypothesis to an analysis of X rays generated by the impact of electrons, and also to the photoelectric effect. Sommerfeld's contribution caused extensive discussions and criticism, especially by Einstein\footnote{Einstein had shortly before the conference criticized in a letter to Besso that Sommerfeld postulated his hypothesis on the role of collision times without any theory. He also discusses some problems with Sommerfeld's hypothesis in the last section of his own report.}. This document shows how desperate the situation was, but it had little influence on further developments.

In this article,  I will concentrate on the foundational aspects of the early quantum theory, the central theme of the first Solvay Congress.\footnote{For a not technically oriented broad discussion that covers several Solvay meetings, including the first one, I refer to \cite{gali}, and references therein.} Therefore, I will almost exclusively describe the lectures by Lorentz, Planck, and Einstein, including their discussions by the plenum. These show the widely different views of the participants. With his conception of a wave-particle duality of the free electromagnetic field, Einstein was completely isolated.
 
\section{Remarks on the report by H. A. Lorentz and its discussion}

The first session was devoted to the report of Lorentz in which he discussed various ways of studying the applicability of the law of equipartition to black body radiation.\footnote{The title  is ``Die Anwendungen des Satzes von der gleichm\"assigen Energieverteilung auf die Strahlung'' (\cite{Euck},  pp. 10-40). [The applications of the theorem of the uniform energy distribution on radiation.]} Lorentz begins by recalling how Lord Rayleigh arrived in 1900 at what is known as the Rayleigh-Jeans formula for the energy distribution $\rho(\nu,T)$ of the black body radiation. Attributing to each mode (oscillator) of the radiation field in a cavity the \textit{equipartition value} $kT$, Rayleigh and Jeans obtained the classic law
\begin{equation}\label{lor1}
\rho(\nu,T)=\frac{8\pi\nu^2}{c^3}kT,
\end{equation}
with its famous difficulties (ultraviolet catastrophe). At the time of the meeting it was known since years that this formula agrees well with experiment in the region with $h\nu/kT\ll 1$. In his revolutionary paper on the light quantum in 1905, Einstein had independently obtained this law as an unavoidable consequence of classical physics (electrodynamics and mechanics). Because of its complete failure in the Wien regime he then concluded that the radiation field has also particle-like properties. (We will come back to this when we discuss Planck's and Einstein's contributions to the congress.)

Much of Lorentz' report is devoted to classical statistical mechanics and a Hamiltonian formulation of the electromagnetic field in interaction with matter (electrons and neutral particles). Within this framework he arrives at the firm conclusion that for the canonical and microcanonical ensembles the average energy of each mode $(\textbf{k},\lambda)$ of the radiation field in thermodynamic equilibrium is $2\times\frac{1}{2}kT$, because the electric and magnetic parts give equal contributions. Therefore, the Rayleigh-Einstein-Jeans law\footnote{Einstein's paper from March 1905 was before Jeans' contribution im May 1905.} for thermodynamic equilibrium is indeed implied by these considerations. At the time, some people (e.g. Jeans) had suggested that the energy exchange between matter and radiation proceeds exceedingly slowly for short wavelengths, so that an equilibrium state is not attained.\footnote{For Jeans the partition law is correct, but `the supposition that the energy of the ether is in equilibrium with that of matter is utterly erroneous in the case of ether vibrations of short wavelength under experimental conditions.'} However, Lorentz discards this possibility in pointing out, that the empirical spectrum deviates from the Rayleigh-Jeans distribution already considerably in the ultrared and visible parts of the spectrum. With numerical examples he also emphasizes that the proportionality of the radiation intensity at a given wavelength with temperature ``is out of the question''.

In addition, Lorentz arrives at the conclusion that the average value of the kinetic energy of a classical electron, with a finite size $a$, must be $\frac{3}{2}kT$, even if radiation with wavelengths $\lambda\leq a$ is cut-off. He summarizes his detailed considerations by stating that a satisfactory radiation formula is definitely outside the classical domain, and that Planck's constant $h$ has to be explained by completely different considerations.

After this he reviews the recent paper by Einstein and Hopf \cite{E_H}\footnote{References to papers that have appeared in the \textit{Collected Papers of Albert Einstein} (CPAE) \cite{Ein0} are always cited by volume and document of CPAE.} on Brownian motion of a moving oscillator in the radiation field, and then applies their approach to the motion of a finite size electron. The general method was developed and applied by Einstein already before \cite{Ein1} , and in a most fruitful manner again several years later (1916). Briefly, Einstein argued as follows. A particle with mass $M$ experiences:\\
$\bullet$ a systematic drag force $-Rv$ that leads in a small time interval $(t,t+\tau)$ to the
momentum change $-Rv\tau$;\\
$\bullet$ an irregular change of momentum $\Delta$ in the time $\tau$, due to fluctuations
of the radiation pressure.
In thermal equilibrium
\[\langle(Mv-Rv\tau+\Delta)^2\rangle=\langle(Mv)^2\rangle. \]
Assuming $\langle v\cdot\Delta\rangle=0$ this implies for sufficiently small $\tau$ the following \textit{fluctuation-dissipation relation}:
\begin{equation}
\langle\Delta^2\rangle=2R\langle M v^2\rangle\tau=2RkT\tau. 
\label{Eq:2}
\end{equation}
 This is one of the magic equations in Einstein's early work on Brownian motion and its implications for the structure of radiation.
 
 In the paper of Einstein and Hopf the material particle (atom, molecule) was idealized as a 1-dimensional harmonic oscillator, that also played a crucial role in Planck's work on black body radiation.
 
 If the free electromagnetic field is described as a classical stochastic field with uncorrelated modes (as a result of translational invariance) and random phases, the rather involved calculations of $R$ and $\langle\Delta^2\rangle$ by Einstein and Hopf led to the following concrete fluctuation-dissipation formula 
 \begin{equation}
 \frac{c^3}{48\pi}\frac{\rho_{\nu}^2}{\nu^2}=\langle E_{kin}\rangle\Bigl(\rho_{\nu}-\frac{1}{3}\nu\frac{\partial\rho_{\nu}}{\partial\nu}\Bigr).
 \label{Eq:3}
 \end{equation}
Here, the left hand side is proportional to the fluctuation $\langle\Delta^2\rangle/\tau$ (due to interferences), and the expression in the bracket on the right is proportional to $R$. This combination comes from a Lorentz transformation for the spectral energy density $\rho_{\nu}$ from the lab system to the rest system of the particle, in which the particle experiences an anisotropic radiation field. The average of the kinetic energy $E_{kin}$ is defined as the time average. Equation (\ref{Eq:3}) holds for general distributions of translational invariant stochastic fields, in particular for equilibrium distributions. This equation is satisfied for the Rayleigh-Jeans distribution, if $\langle E_{kin}\rangle=\frac{1}{2}kT$.

Lorentz expresses some doubts as to whether the high-frequency oscillations of the Planck oscillator can be described by classical mechanics plus electrodynamics. Therefore, he considers instead a classical finite size electron model. Without derivation he presents his key formula for the 1-dimensional kinetic energy $E_{kin}$ in terms of the distribution function:
\begin{equation}
\langle E_{kin}\rangle=\frac{c^3}{64\pi}\frac{\int_{0}^{\infty}(\rho_{\nu}^2/\nu^2)\;d\nu}{\int_{0}^{\infty}\rho_{\nu}\;d\nu}.
\label{Eq:4}
\end{equation} 
Lorentz first applies (\ref{Eq:4}) for a Rayleigh-Jeans distribution that is sharply cut-off at some limiting frequency $K$, and obtains independent of $K$
\begin{equation}
\langle E_{kin}\rangle=\frac{1}{4}\times\frac{1}{2}kT.
\label{Eq:5}
\end{equation}
His comment: ``I had hoped (...) to find $\langle E_{kin}\rangle=\frac{1}{2}kT$.'' (I will come back to this unexpected factor 1/4.)

\subsection{A great mystery}

A possible equilibrium distribution that satisfies Wien's displacement law (equivalent to the adiabatic invariance of $\rho(\nu,T)/\nu^3$) must be of the general form
\begin{equation}
\rho(\nu,T)=\nu^3\varphi(T/\nu).
\label{Eq:6}
\end{equation}
Then equation (\ref{Eq:4}) can be written as
\begin{equation}
\langle E_{kin}\rangle=\frac{c^3}{64\pi} T\; \frac{\int_{0}^{\infty} (\varphi(x)^2/x^6)\;dx}{\int_{0}^{\infty} (\varphi(x)/x^5)\;dx}.
\label{Eq:7}
\end{equation}
Note that this is \textit{linear} in $T$. Inserting the Planck distribution, Lorentz obtains numerically 
\begin{equation}
\langle E_{kin}\rangle\simeq \frac{1}{25}\times \frac{1}{2}kT. 
\label{Eq:8}
\end{equation}
This dramatic failure was already the main point of Einstein and Hopf. At the end of their paper they wrote\footnote{``additional kinds of momentum fluctuations are discernible (...) which, in the case of short-wave radiation of low density, enormously overwhelm those obtained from the theory''.}: Bei kurzwelliger Strahlung geringer Dichte treten Impulsschwankungen anderer Art auf, welche die klassischen ``ungeheuer \"uberwiegen''.

Lorentz asks whether it might be possible to overcome the factor 1/25 in (\ref{Eq:8}) by choosing a distribution function with a sufficiently sharp maximum such that the numerator in (\ref{Eq:7}) becomes larger, without changing the denominator. But that seems to be unlikely, he says, without deviating too much from the empirically successful Planck distribution. 

Therefore, he concludes that the energy of electrons in interaction with black body radiation can not reach the value $\frac{3}{2}kT$, \textit{if only interference oscillations are responsible}. There must be other momentum fluctuations of particles (oscillators, electrons) due to irregularities in the radiation field, which are much larger than the classical ones for small density of the radiation energy. In his concise article ``Einstein's contributions to quantum theory'' Pauli wrote about the paper of Einstein and Hopf \cite{pauli}: ``This result was disappointing for those who still had the vain hope of deriving Planck's radiation formula by merely changing the statistical assumption rather than by a fundamental break with the classical ideas regarding the elementary micro phenomena themselves.'' 

\subsection{The strange factor 1/4 and an absurd discussion}

Much of the discussion on Lorentz' report centred around the disturbing factor 1/4 in (\ref{Eq:5}). First, Langevin came up with a different calculation of $\langle E_{kin}\rangle$, based on an ordinary differential equation and not considering fluctuations, that produced $\frac{1}{2}kT$, but Lorentz criticized this, because Langevin ignored magnetic forces, and he suggested that the difference might be a result of this omission. After that Poincar\'{e} made an irrelevant remark on radiation damping. Then Planck doubted that Einstein's basic formula (\ref{Eq:2}) can be applied to the motion of electrons in the field of black body radiation, and he also proposed another way of computing $\langle E_{kin}\rangle$ that gave the value $\frac{1}{2}kT$. This was, however, completely unphysical, as Lorentz commented. By that Planck effectively reduced his calculation to that of Langevin. At some point Einstein concluded that ``neither the consideration of Mr. Langevin nor that of Mr. Planck solves the problem, in my opinion.'' Einstein gave, however, no explanation of the factor 1/4, but he was presumably also puzzled.

Hundred years after this discussion I discovered how the wrong factor 1/4 was produced. The reason is entirely trivial: Integrate the fluctuation-dissipation relation (\ref{Eq:3}) of Einstein and Hopf over frequencies. If $\rho_{\nu}$ is smooth and vanishes for $\nu\rightarrow\infty$, then one can perform a partial integration in the last term, and obtains exactly the formula (\ref{Eq:4}) of Lorentz. However, for the cut-off Rayleigh-Jeans distribution one gets a boundary term whose inclusion compensates the factor 1/4. Presumably, nobody else ever noticed this.

Well ! ``If wise men did not err fools would despair.'' (W. Goethe)

\subsubsection{Remarks on the missing fluctuations}

The missing fluctuations in the Einstein-Hopf fluctuation-dissipation relation were found by Einstein five years after the Solvay congress, when he studied in one of his great papers again the Brownian motion of an atom or molecule in the radiation field. The first part of his famous paper \cite{ein16} is known to all physicists, because it contains a purely quantum derivation of Planck's distribution, and also the theoretical foundations of the laser. He thereby introduced the hitherto unknown process of
induced emission\footnote{Einstein's derivation shows that without assuming a non-zero probability for induced emission one would necessarily arrive at Wien's instead of Planck's radiation law.}, next to the familiar ones of spontaneous emission and induced absorption. 

The second part of the paper is, however, much less known, but was regarded by Einstein to be the more important. There he shows that in every elementary process of radiation, and in particular in spontaneous emission, an amount $h\nu/c$ of momentum is emitted in a random direction and that the atomic system suffers a corresponding recoil in the opposite direction. This recoil was first experimentally confirmed in 1933 by R.\,Frisch \cite{frisch}, when he showed that a long and narrow beam of excited sodium atoms widens up after spontaneous emissions have taken place. Using his rules for emission and absorption processes in terms of the coefficients $A$ and $B$ (with two relations among them), Einstein now obtained instead of (\ref{Eq:3}) the following extension of the fluctuation-dissipation relation:
\begin{equation}
 \frac{c^3}{16\pi\nu^2} \Bigl[\rho_{\nu}^2 + \frac{8\pi\nu^2}{c^3} h\nu\rho_{\nu}\Bigr]=\langle E_{kin}\rangle\Bigl(\rho_{\nu}-\frac{1}{3}\nu\frac{\partial\rho_{\nu}}{\partial\nu}\Bigr)
 \label{Eq:9}
 \end{equation}
($\langle E_{kin}\rangle$ is now the kinetic energy of the atom). The second term on the left gives the quantum part of the momentum fluctuations, and is linear in $\rho_{\nu}$. This relation is identically satisfied for the Planck distribution if 
$\langle E_{kin}\rangle=\frac{3}{2}kT$. It overcomes the small factor $\simeq 1/25$ stressed by Lorentz in his report. Conversely, relation (\ref{Eq:9}), considered as a differential equation for $\rho_{\nu}$, implies `almost' the Planck distribution.

The quantum term in (\ref{Eq:9}) was not a surprise to Einstein. He had studied momentum fluctuations already in 1909 \cite{Ein1}, in considering the Brownian motion of a mirror which perfectly reflects radiation in a small frequency interval, but transmits for all other frequencies. The final result he commented as follows:
\begin{quote}
The close connection between this relation and the one derived in the last section for the energy fluctuation is immediately obvious, and exactly analogous considerations can be applied to it. Again, according to the current theory, the expression would be reduced to the first term (fluctuations due to interference). If the second term alone were present, the fluctuations of the radiation pressure could be completely explained by the assumption that the radiation consists of independently moving, not too extended complexes of energy $h\nu$.
\end{quote}
Nobody would follow Einstein's conception for many years to come. The theme will be taken up again when we come to his talk at the Solvay meeting.

%%%%%%%%%%%%%%%%%%%%%%%%%%%%%%%%%%%%%%%%%%%%%%%%%%%%%%%%%%%%%%%%%%%%%

\section{Planck's report and its plenary discussion}

In his written report (\cite{Euck}, pp. 77-94) Planck reviews the various theoretical attempts in trying to understand the empirically successful Planck distribution of the black body radiation. As he explicitly states, he does not follow the historical sequence, but that his presentation is guided by the contents. For this reason it may be necessary to briefly recall Planck's original derivation in December 1900 \cite{planck1}.

\subsection{Planck's discovery of energy quanta}

Sometime before the discovery of his radiation law, Planck had established a relation between the black body distribution function $\rho(\nu,T)$ and the equilibrium energy $\bar{E}_{\nu}$ of a linear oscillator with frequency $\nu$ in thermodynamic equilibrium with the radiation field. We emphasize that this was obtained on the basis of \textit{classical} physics (mechanics and electrodynamics) and reads
\begin{equation}\label{P1}
\rho(\nu,T)=\frac{8\pi\nu^2}{c^3}\bar{E}_{\nu}.
\end{equation}
(For a textbook derivation, see \cite{Som}.) At this point we argue somewhat backwards. From the relation (\ref{P1}) and Planck's distribution law
\begin{equation}\label{P2}
\rho(\nu,T)=\frac{8\pi\nu^2}{c^3}\frac{h\nu}{e^{h\nu/kT} - 1},
\end{equation}
originally guessed by Planck on the basis of new measurements, we obtain
\begin{equation}\label{P3}
\bar{E}_{\nu}=\frac{h\nu}{e^{h\nu/kT} - 1}.
\end{equation}
Using now the thermodynamic relation $dS=\frac{1}{T} dE,\:(E\equiv \bar{E}_{\nu})$ for the entropy of the oscillator, we obtain from (\ref{P3})
\[dS=\frac{k}{h\nu} \ln \left (1+\frac{h\nu}{E}\right )dE, \]
thus
\begin{equation}\label{P4}
S=k\left [\left (1+\frac{E}{h\nu}\right )\ln \left (1+\frac{E}{h\nu}\right )-\frac{E}{h\nu}\ln \frac{E}{h\nu}\right ].
\end{equation}
Conversely, once we understand this expression for the entropy, Planck's distribution follows, provided we accept the relation (\ref{P1}) that was derived with classical physics. (We know that equation (10) indeed holds in Dirac's radiation theory.)

Now, we present Planck's first derivation of (\ref{P4}), about which Einstein wrote in his `Autobiographical Notes' \cite{Ein2} in 1949: ``the imperfections of which remained at first hidden, which latter fact was most fortunate for the development of physics''. Planck applied Boltzmann's  principle to a system which consists of a large number $N$ of linear oscillators, with the same frequency $\nu$. Let $E_N$ be the total energy  and $S_N$ the total entropy of the system. According to Boltzmann $S_N=k \ln W_N$, where $W_N$ is the `probability' of the macro-state. In order to express the number of micro-states belonging to the macro-state, Planck divided the total energy $E_N$ into a large number $P$ of energy elements $\varepsilon: \,E_N=P\varepsilon$, and defined $W_N$ to be the number of ways in which the $P$\textit{ indistinguishable} energy elements can be distributed over $N$ \textit{distinguishable} oscillators. A well-known combinatoric argument gives 
\begin{equation}\label{P5}
W_N=\frac{(N-1+P)!}{P! (N-1)!}.
\end{equation}
If this is accepted, we obtain for the entropy with the Stirling approximation
\begin{equation}\label{P6}
S_N\simeq k[(N+P)\ln (N+P)-N\ln N-P\ln P].
\end{equation}
For $S=S_N/N$ and $E=E_N/N =\varepsilon P/N$ this becomes equation (\ref{P4}) if $\varepsilon=h\nu$.

But how can this counting of the number of micro-states (that led to the Bose-Einstein distribution of 1925) be justified? What is the physical meaning behind Planck's counting of complexions? Much later (in 1931) Planck referred to it as an ``act of desperation (...). I had to obtain a positive result, under any circumstances and whatever cost'' \cite{planck2}.

It should be stressed at this point that Planck did \textit{not}, as is often stated, quantize
the energy of a material oscillator per se. As we have seen, what he was actually doing
in his decisive calculation of the entropy of a harmonic oscillator
was to assume that the \emph{total} energy of a large number of
oscillators is made up of \emph{finite} energy elements of equal
magnitude $h\nu$. He did not propose that the energies of single material
oscillators are physically quantized.\footnote{\label{foot:Planck}
In 1911 Planck even formulated a `new radiation hypothesis', in which
quantization only applies to the process of light emission but not to
that of light absorption \cite{planck1911}. Planck's explicitly stated
motivation for this was to avoid an effective quantization of oscillator
energies as a \emph{result} of quantization of all interaction energies.
It is amusing to note that this new hypothesis led Planck to a
modification of his radiation law, which consisted in the addition of
the temperature-independent term $h\nu/2$ to the energy of each
oscillator, thus corresponding to the oscillator's energy at zero
temperature. This seems to be the first appearance of what soon
became known as `zero-point energy'.} Rather, the energy elements
$h\nu$ were introduced as a formal counting device that could at
the end of the calculation not be set to zero, for, otherwise, the
entropy would diverge. These energy elements are at the beginning of the quantum revolution.\footnote{There is the story (which I heard from M. Fierz) that Planck used the letter $h$ for his constant, because since the times of Cauchy the differential quotient of a function was defined as the limit of a difference quotient, in which the increment of the argument -- universally denoted by $h$ in all text books since then -- is approaching 0.}

It was Einstein in 1906 who interpreted Planck's result as follows \cite{Ein3}:
\begin{quote}
Hence, we must view the following proposition as the
basis underlying Planck's theory of radiation: The energy of an
elementary resonator can only assume values that are integral
multiples of $h\nu$; by emission and absorption, the energy of
a resonator changes by jumps of integral multiples of $h\nu$.
\end{quote}
His line of thought will be discussed in the next subsection.

In his report, Planck repeated his considerations of 1900 that led to Eq. (\ref{P4}), and thus to (\ref{P3}). Shortly before the Solvay Congress, L. Natanson had emphasized the arbitrariness of Planck's counting procedure that led to Eq. (\ref{P5}) \cite{Nat}. (Counting partitions of \textit{indistinguishable} objects -- the energy elements -- was hardly in the spirit of Boltzmann.) This criticism was accepted by Planck in his report, but he now states that the counting procedure is actually unique\footnote{In a footnote Planck writes: ``This calculation is completely unambivalent and in particular no longer contains the indefiniteness about which L. Natanson has recently spoken with justification.''  [``Diese Berechnung ist vollkommen eindeutig und enth\"{a}lt insbesondere nichts mehr von der Unbestimmtheit, welche L. Natanson neuerdings mit Recht zur Sprache gebracht hat.'']}, if Einstein's interpretation, cited above, is adopted.\footnote{It is somewhat disturbing that Planck does not refer to Einstein at this point.} Then it is at least natural to consider the energy states $h\nu n, \,n=0,1,2, ... $ as cases, whence, in obvious notation,
\[W_N= \sharp \left \lbrace\left.  (n_1,n_2, ...\,, n_N)\right \vert \sum_{i=1}^{N}  n_i =P,\: n_i \in \mathbb{N} _0\right \rbrace. \]
This agrees, of course, with the number (\ref{P5}).\footnote{In passing we note that this number is equal to the degeneracy of an energy state with energy $h\nu\,P$ of $N$ quantum mechanical oscillators. For this reason, Planck got the same entropy that is implied by the microcanonical ensemble of quantum statistics.} Einstein's main criticism of Planck's procedure is quoted below (at the beginning of Sect. 3.3).

How did Planck arrive at this interpretation of the energy elements $h\nu$ after his pioneering paper? This is discussed extensively earlier in his report. Briefly, he arrived close to the quantization rule
\begin{equation}\label{P7}
\oint p\,dq=n\, h, \quad n=0,1,2,... \, ,
\end{equation}
for systems with one degree of freedom. Certainly, his treatment of the harmonic oscillator, that postulates indivisible elementary regions in the $q-p$ -- manifold, can be interpreted as an application of this quantization rule: Using
\begin{equation}\label{P8}
\int_{{H\leq E}} dp \,dq =\oint_{{H=E}}p \,dq= n\, h
\end{equation}
for the Hamilton function of the harmonic oscillator, $H=\frac{1}{2m}p^2 +\frac{1}{2}m\omega^2 q^2$, implies immediately that $E=nh\nu$. Planck emphasizes that the phase space region over which one has to integrate on the left is bounded be an ellipse, and that the surface between $E$ and $E+\varepsilon$ is equal to the ``elementary action'' $h$. He concludes with: ``Bei einem Oszillator von bestimmter Eigenperiode $\nu$ existieren also bestimmte Energieelemente $\varepsilon=h\nu$ insofern, als f\"{u}r die Wahrscheinlichkeit einer bestimmten Energiegr\"{o}sse lediglich die Anzahl der Energieelemente massgebend ist, die sie umfasst.'' [For an oscillator of given frequency $\nu$ there thus exist definite energy elements $h\nu$ in so far, as for the probability of a given magnitude of energy only the number of energy elements matters, that it contains.]

It is remarkable that it still took some time until even the most outstanding people correctly applied the quantization rule (\ref{P7}) to periodic systems with one degree of freedom. After several wrong attempts, in particular by Einstein and Stern, the correct result for the rotator with fixed axis was only given in 1913 by Ehrenfest. (For this strange story, see \cite{Hund}, p. 36.) Einstein's report ends, as we shall see, with remarks on rotating molecules. The generalization to higher dimensions was at that time completely unknown. An interesting difficulty was pointed out by Einstein in a remark to Nernst's lecture (see CPAE, Vol. 3, Doc. 25).

\subsection{Einstein's interpretation of Planck's work}

In 1906 Einstein re-analysed Planck's approach, and asked how the classical microcanonical ensemble for  large number of oscillators should be modified to obtain formula  (\ref{P3}) \cite{Ein3}. We have already quoted his conclusion. Soon afterwards, he studied the same question in the context of the canonical ensemble \cite{Ein4}. Planck cites this work, and describes it as a second main way to obtain the mean energy (\ref{P3}) of a harmonic oscillator. Since Einstein also repeated this approach in his contribution, we discuss it here.

Classically, the canonical partition sum is 
\begin{equation}
Z=\int e^{-\beta H}\,d\Gamma, \; \beta=1/kT,
\end{equation}
where $d\Gamma$ is the Liouville measure on phase space. The mean energy is given by $\langle H\rangle=-\frac{\partial}{\partial\beta}\ln Z$. Einstein rewrites $Z$  as 
\begin{equation}
Z=\int e^{-\beta E}\omega(E)\,dE,  \quad \mbox{where} \quad \omega(E)= \int \delta(H-E)\,d\Gamma
\end{equation}
is the volume of the energy surface $\lbrace H=E\rbrace$. For a harmonic oscillator one obtains $\omega(E)=\mbox{const.},\, Z\propto 1/\beta$, and thus the standard equipartition result $\langle H\rangle=kT$. 
In quantum theory, Einstein performs the substitution
\begin{equation}\label{ein2}
\omega(E) \longrightarrow \mbox{const} \times\sum_{n=0}^{\infty} \delta(E-n\varepsilon), \quad\varepsilon= h\nu,
\end{equation}
and obtains the crucial formula (\ref{P3}). This is then used in the same paper for his theory for the specific heat of solids, to which we will come when discussing Einstein's contribution at the Solvay Congress.

After this Planck discusses two other ways of obtaining the same results, one of which is more in the spirit of Boltzmann's statistical conception of entropy. Then he comes to a critical analysis of the current situation. His views are fundamentally different from those of Einstein, because he wanted to maintain by all means Maxwell's theory in vacuum, and apply the quantum hypothesis only to matter that interacts with radiation. As all the other colleagues (with the exception of J. Stark), he was for many years against Einstein's light quanta and his particle-wave duality of radiation, to which we will come later in the section on Einstein' report and its discussion. Already in 1907 Planck wrote to Einstein \cite{planck3}:
\begin{quote}
I am not seeking the meaning of the quantum of action in the vacuum but rather in places where absorption and emission occur, and I assume what happens in the vacuum is rigorously described by Maxwell's equations.
\end{quote}

\subsection{Discussion of Planck's report}

The extensive discussion of Planck's written report, covering 14 pages in the proceedings, was initiated by Einstein. Because of its importance, we quote it in extenso\footnote{This is a summary of Einstein's critique he had earlier presented in \cite{Ein5}, pp. 187-188.}:

\begin{quote}
What I find strange about the way Mr. Planck applies Boltzmann's equation is that he introduces a state probability $W$ without giving this quantity a physical definition. If one  proceeds in such a way, then, to begin with, Boltzmann's equation does not have a physical meaning. The circumstance that $W$ is equated to the number of complexions belonging to a state does not change anything here; for there is no indication of what is supposed to be meant by the statement that two complexions are equally probable. Even if it were possible to define the complexions  in such a manner that the $S$ obtained from Boltzmann's equation agrees with experience, it seems to me that with this conception of Boltzmann's principle it is not possible to draw any conclusions about the admissibility of any fundamental theory whatsoever on the basis of the empirically known thermodynamic properties of a system.
\end{quote}

The interesting part of the discussion centred on the question of whether, in our terminology, the electromagnetic field in vacuum (``the ether'') could remain classical or had also to be quantized. The question was first taken up by Jeans. In his answer Planck expressed again his view that the quantum of action plays only a role in emission and absorption processes. In this connection Langevin mentioned a paper of Debye from 1910 \cite{debye}, that was also cited in passing by Planck in his report. For me it is astonishing, that Debye's approach to the Planck distribution did not get more attention. Debye quantized directly the oscillators of the electromagnetic field. Later, in 1913, he applied the same method in his classical theory of the specific heat, every student of physics nowadays learns about. I believe it is appropriate to go briefly through the main steps of Debye's study.
\[\qquad \qquad\qquad\ast\ast\ast \qquad \qquad\qquad\ast\ast\ast \qquad \qquad \ast\ast\ast \qquad\qquad \qquad \]
\textit{Digression}. Debye represents the spectral energy of the radiation field in a cube of volume $V$ as
\begin{equation}\label{debye1}
U_\nu\,d\nu =V \frac{8\pi\nu^2}{c^3} h\nu f(\nu) d\nu \equiv N(\nu) f  h\nu d\nu, \quad  N(\nu):= V \frac{8\pi\nu^2}{c^3},
\end{equation}
where $f(\nu)$ is a general frequency distribution function for the energy quanta $h\nu$. Following Planck's counting, Eq. (\ref{P5}), he associates to the number of oscillators $N(\nu)\Delta\nu$ in a small frequency interval $\Delta\nu$ the number (of micro-states)
\begin{equation}\label{debye2}
w=\frac{(N\Delta\nu+Nf\Delta\nu)!}{(N\Delta\nu)!(Nf\Delta\nu)!}.
\end{equation}
Instead of (\ref{P6}), Debye obtains for the entropy
\begin{equation}\label{debye3}
S/k=V \frac{8\pi}{c^3}\int_{0}^{\infty} \lbrace (1+f) \ln (1+f)-f \ln f \rbrace \nu^2\,d\nu
\end{equation}
for an \textit{arbitrary} state of radiation with a given spectral energy (\ref{debye1}). In thermodynamic equilibrium the entropy reaches a maximum for a given total energy
\begin{equation}
U=V\frac{8\pi h}{c^3}\int \nu^3 f(\nu)\,d\nu.
\end{equation}
With the method of Lagrange multipliers, and using the thermodynamic relation $dS/dU=1/T$ to determine the Lagrange multiplier, Debye obtains the distribution function
\begin{equation}
f=\frac{1}{e^{h\nu/kT}-1},
\end{equation}
and Planck's expression  (\ref{P3}) for the entropy.
\[\qquad \qquad\qquad\ast\ast\ast \qquad \qquad\qquad\ast\ast\ast \qquad \qquad \ast\ast\ast \qquad\qquad \qquad \]

Debye concludes in stating that his derivation does not really prove whether the existence of elementary quanta is a property of the ``ether'', as he prefers for the time being, or a property of matter. An unquestionable advantage is, however, that Debye did not have to assume Planck's relation (\ref{P1}), which was derived with classical physics. Einstein's criticism of Planck's counting of micro-states applies, of course, also to Debye's work. Fortunately, it became very fruitful in his theory of the specific heat of solids at low temperatures, published shortly after the Solvay Congress \cite{debye2}.\footnote{Debye gave a talk on his theory in March 1912 at a meeting of the Swiss Physical Society, and published a brief version in the \textit{Arch. de Gen\`{e}ve} shortly afterwards. During this time, after Einstein had left for Prag, Debye was Prof. at the at the University of Z\"{u}rich.}

At some point Einstein injected the remark: ``Objections have often been raised against the application of statistical methods to radiation. But I do not see any reason why these methods should be excluded here.''

%%%%%%%%%%%%%%%%%%%%%%%%%%%%%%%%%%%%%%%%

\section{Einstein's report: "\textit{On the Present State of the Problem of Specific Heats}"}

The first section of Einstein's report is entitled: ``The connection between specific heats and the radiation formula.'' Here one finds equations which appeared already in Planck's contribution, but the prose is different in important ways. He starts with: ``Let thermal radiation, an ideal gas, and oscillators of the kind indicated be enclosed in a volume bounded by perfectly reflecting walls. By virtue of their electric charges, the oscillators must emit radiation and continually receive new momentum from the radiation field. On the other hand, the material point of the individual oscillator collides with gas molecules and in this way exchanges energy with the gas. The oscillators thus bring about an energy exchange between the gas and radiation, and the energy distribution of the system in the state of statistical equilibrium is completely determined by the total energy, if we assume that oscillators of all frequencies are present.'' 

After that he recalls Planck's relation (\ref{P1}), which was derived on the basis of classical mechanics and  Maxwell's theory. Then Einstein continues with: ``On the other hand, statistical mechanics implies the following: If the volume contains only gas and oscillators (without charge), there is a relation between the temperature $T$ and the mean energy $\bar{E}_{\nu}$ of the three-dimensional oscillator of the form 
\begin{equation}\label{ein1}
\bar{E}_{\nu}=3kT.
\end{equation}

But if the oscillators interact simultaneously with the radiation and the gas, then (\ref{P1}) and (\ref{ein1}) must be satisfied simultaneously if they hold individually in the special cases discussed; for if one of these equations were not satisfied, this would result in a transport of energy, whether between radiation and resonators, or between gas and resonators."

The two equations imply the radiation distribution (\ref{lor1}), that should actually be called the Rayleigh-Einstein-Jeans law (for justification, see \cite{Pais}, Sect. 19b). Einstein re\-peats, what he had stated already in 1905: ``This is the only radiation equation that is simultaneously in agreement with our mechanics and electrodynamics.'' In view of the fact that it does not correspond to reality, he goes on as follows (which again illustrates the wonderful clarity in Einstein's reasoning):
\begin{quote}
Faced with this failure of our theories to conform to reality, Planck proceeds in the following fashion: He rejects (\ref{ein1}), and thereby a foundation in mechanics, but keeps (\ref{P1}), even though mechanics has been applied in the derivation of (\ref{P1}) as well. He obtains his theory of radiation by replacing (\ref{ein1}) by a relation in whose derivation he introduced, for the first time, the quantum hypothesis. However, for what follows, we need neither (\ref{ein1}) nor a corresponding relation, but only equation (\ref{P1}). Even if we abandon (\ref{ein1}), we must adhere to the proposition that (\ref{P1}) is valid not only when the oscillator is influenced by the radiation alone, but also when molecules of a gas having the same temperature collide with the oscillator. Because if these molecules were to alter the mean energy of the oscillator, then more radiation would be emitted by the oscillator than it absorbs, or vice versa. Equation (\ref{P1}) also remains valid when the energy variations in the resonator are preponderantly determined by the interaction between the oscillator and the gas; it is certainly therefore also valid in the total absence of an interaction with radiation, for example, when the oscillators have no charge whatsoever. The equation is also valid if the body interacting with the oscillator is not an ideal gas but any other kind of body, as long as the oscillator vibrates approximately monochromatically.
\end{quote}
Using now Planck's radiation formula ``confirmed to the highest degree of approximation'', the formula (\ref{P3}), discussed  before by Planck follows. We repeat it here for an oscillator with three degrees of freedom
\begin{equation}
\bar{E}_{\nu}=3\frac{h\nu}{e^{h\nu/kT} - 1}.
\end{equation}
We recall that Einstein had derived this formula in his original paper from 1907 on the specific heat from a quantum version of the canonical ensemble.

\subsection{Einstein's model for the specific heat of a solid}

Now Einstein assumes ``that one gram-atom of a  solid consists of $N_A$ (= Avogadro number) such approximately monochromatic oscillators'', he obtains by differentiation its specific heat ($R=kN_A$)
\begin{equation}\label{ein3}
c_v=3R\frac{(h\nu/kT)^2e^{h\nu/kT}}{\left (e^{h\nu/kT} -1\right )^2}.
\end{equation}
Einstein demonstrates with an accompanying figure that this simple theory agrees remarkably well with measurements by Nernst. His comment: ``Even though systematic differences between the observed and the theoretical values do exist, the agreement is nevertheless astonishing, if one takes into account that each individual curve is completely  determined by a single parameter $\nu$, namely the proper frequency of the atom in question.'' Referring to work by himself and others, in particular of Madelung, Einstein states:
``In my opinion, the cause for this deviation  must be sought in the fact that thermal oscillations of the atoms deviate markedly from monochromatic oscillations, and therefore do not actually have a definite frequency but rather a range of frequencies.'' I do not elaborate more on this, and also not on a modified formula by Nernst and Lindemann, because soon afterwards Debye \cite{debye2}, and independently Born and K\'{a}rm\'{a}n \cite{born1}, \cite{born2} developed a satisfactory theory in this direction. Later, we will add some remarks on this.

Einstein's work on specific heat was less profound than his investigations on radiation, but it played an important role because it involved the quantum hypothesis in another domain, in which people like Nernst were experimentally active.\footnote{Nernst presented at the Solvay Congress a detailed report on his measurements. The fitting formula  proposed by him and Lindemann had no theoretical basis, but it later turned out that it agrees numerically surprisingly well with Debye's result over a wide range of temperatures ($0<\Theta/T <4$). 

As noted before, F. A. Lindemann was, together with R. Goldschmidt and M. de Broglie, appointed as secretary of the conference. He was the youngest attendee. Frederick Lindemann did his doctoral thesis with Nernst. Much later he became scientific advisor to Winston Churchill. He was made \textit{Lord Cherwill} in 1941 and Viscount Cherwill in 1956. To physicist he is, for instance, known for the \textit{Lindemann melting criterion}.} 

In \S2 of his report, Einstein turns ``to the highly important but, unfortunately, mainly unsolved question: How is mechanics to be reformulated so that it does justice to the radiation formula as well as the thermal properties of matter?'' First he presents his derivation of the average energy of an oscillator based on (\ref{ein2}), i.e., in replacing the energy integral of the classical theory by a sum of oscillator energies $n h\nu$. Einstein's comment on this is illuminating: ``Simple as this hypothesis is, and simple as it is to arrive at Planck's formula with its aid, it contents strike us as counter-intuitive and outlandish on closer inspection. Let us consider a diamond atom at 73 K: What can be said about the oscillation of the atom on the basis of Planck's hypothesis? If, with Nernst, we set $\nu=27.3\times 10^{12}$ Hz, we obtain from the oscillator formula $\bar{E}/h\nu=e^{-18.6}$. (...) Only one of $10^8$ atoms oscillates at any given moment, while the others are completely at rest. No matter how firm one's conviction that our current mechanics is not applicable to such motions, such a picture strikes one as extremely strange.''

\subsection{Energy fluctuations of a solid}

After some supplementary remarks, Einstein considers fluctuations of the energy of a solid body. As in previous work, he derives from Boltzmann's equation $S=k \ln W$ + const. the statistical probability (Einstein's expression) $W$. Expanding $S(E)$ about the mean value of the energy $\bar{E}$ of the body, he obtains for the mean square deviation of the energy from the mean value the general result of what we now call Einstein's fluctuation theory:
\begin{equation}\label{ein4}
\overline{(\Delta E)^2} =- \frac{k}{\partial^2 S/\partial E^2}.
\end{equation}
Here, $\partial^2 S/\partial E^2$ has to be evaluated for the equilibrium entropy at $\bar{E}$, keeping the volume fixed. In making also use of the thermodynamic relation $\partial^2 S/\partial E^2=-1/c_v T^2$, Einstein arrives at
\begin{equation}\label{ein4'}
\overline{(\Delta E)^2} =k c_v T^2.
\end{equation}

He emphasizes that this formula is completely general. (He had obtained it in classical statistical mechanics already in 1904.) Inserting his result (\ref{ein3}) for the specific heat for $n$ gram-atoms having the frequency $\nu$, and eliminating $T$ with the aid of
\begin{equation}
\bar{E}=3n N_A\frac{h\nu}{e^{h\nu/kT} - 1},
\end{equation}
Einstein obtains the important fluctuation formula
\begin{equation}\label{ein5}
\frac{\overline{(\Delta E)^2}}{\bar{E}^2}=\frac{h\nu}{\bar{E}} +\frac{1}{3n N_A}=\frac{1}{Z_q}+\frac{1}{Z_f},
\end{equation}
``where $Z_q$ denotes the average number of Planck's `quanta' found in the body, and $Z_f=3n N_A$ the total number of degrees of freedom of all the atoms of the system taken together.''\footnote{We note that for $N$ quantum mechanical oscillators with equal frequencies $\nu$, one readily obtains for the expectation value of the Hamiltonian $\langle H\rangle=N h\nu \bar{n},\,\bar{n}$ = mean occupation number of one oscillator, and for the variance $\sigma^2(H)$
\[\frac{\sigma^2(H)}{\langle H\rangle^2}=\frac{1}{N}\Bigl(1+\frac{1}{\bar{n}}\Bigr).\]
}

As in his previous work on fluctuations of the radiation field, which gave him enormously important insight, Einstein finds again two terms of completely different origin. We quote:
\begin{quote}
The relative fluctuation corresponding to the second term, which is the only fluctuation according to our mechanics, results from the fact that the number of degrees of freedom of the body is finite; it is independent of the magnitude of the energy content. But the relative fluctuation corresponding to the first term has nothing to do with how many degrees of freedom the body has. This fluctuation depends solely on the proper frequency, and the magnitude of the mean energy, and vanishes when this energy is very large. The magnitude of this fluctuation shows an exact agreement with the quantum hypothesis, according to which energy consists of quanta  of magnitude $h\nu$, which change their location independently of each other; indeed, neglecting the second term, the equation can be written in the form
\begin{equation}\label{ein6}
\sqrt{\frac{\overline{(\Delta E)^2}}{\bar{E}^2}}=\frac{1}{\sqrt{Z_q}}.
\end{equation}
\end{quote}

Einstein adds lots of comments and difficult questions to this, which are presumably related to his efforts of developing, what he had called in his famous Salzburg lecture of 1909 \cite{Ein6} `` a kind of fusion of the wave and emission theories of light''. We include here only one of his remarks. Einstein asks: ``Does the fluctuation equation just derived exhaust the thermodynamic content of Planck's radiation formula or of Planck's equation for the oscillator (\ref{P3})? It can easily be seen that this is indeed the case.''  Just substitute in the general fluctuation formula (\ref{ein4'}), i.e. 
\begin{equation}
\overline{(\Delta E)^2} =k T^2 \frac{d\bar{E}}{dT},
\end{equation}
for the left hand side the formula (\ref{ein5}), we obtain (\ref{P3}) by integration. Thus, Einstein concludes, ``a mechanics that would lead to the equation we derived from the energy fluctuations of an ideal solid would also have to lead to Planck's oscillator formula.'' In a third section on ``the quantum hypothesis and the general character of the related experiments'', Einstein begins with:
\begin{quote}
The positive results produced by the investigations described in the last section can be summarized as follows: When a body absorbs  or emits thermal energy by a quasi-periodical mechanism, the statistical properties of the mechanism are such as they would be if the energy were propagated in whole quanta of the magnitude $h\nu$. Though we have little insight into the details of the mechanism by which nature produces this property of these processes, we must expect all the same that the disappearance of such an energy of a periodic character is accompanied by the generation of packets of energy in the form of discrete quanta of magnitude $h\nu$, and second, that energy in discrete quanta of magnitude $h\nu$ must be available, so that energy of a periodic character in the frequency region $\nu$ may be produced. (...) These discontinuities, which we find so off-putting in Planck's theory, seem really to exist in nature.

The difficulties which stand in the way of formulating a satisfactory theory of these fundamental processes seem insurmountable at this time. From where does an electron in a piece of metal that is struck by Roentgen rays take the great kinetic energy we are seeing in secondary cathode rays? After all, the field of the Roentgen rays impinges on all of the metal; why does only a small portion of electrons attain the velocity of those cathode rays? How is it that the absorbed energy shows up only in relatively exceedingly few places? What distinguishes these places from other places? These and many other questions are being asked in vain.
\end{quote}

\subsection{Rotational energies of two-atomic molecules}
We conclude with a remark on Einstein's final topic, the \textit{average rotational energy of two-atomic molecules}. It is somewhat astonishing that he and others were at the time not able to arrive at the correct expression for the quantized energies of a rotator with a fixed axis. From (\ref{P7}) this would have been obvious; this quantization rule leads immediately to the discrete energies 
\begin{equation}
E_n=\frac{h^2}{8\pi^2 I}\,n^2, \; n=0,1,2, ...\,, 
\end{equation}
where $I$ is the moment of inertia. The mean energy of the rotator is then determined by the corresponding canonical partition sum, as previously for a harmonic oscillator. However, this is not what Einstein does. We have already indicated that it took quite a while until people treated the specific heat of rotating molecules correctly. The discussion after Einstein's lecture on this shows how confusing the situation was. The correct result by Ehrenfest in 1913 \cite{Ehren} for a rotator with a fixed axis is presented in an appendix to the proceedings by Arnold Eucken.\footnote{Quantization of rotators with two degrees of freedom in the framework of the old quantum theory (Bohr-Sommerfeld quantization rules) began only in 1915 by M. Planck, F. Reiche, E.C. Kemble, N. Bohr, and others.} In addition, good measurements at low temperatures by him, in particular for molecular hydrogen, became only available in 1912. (The German edition of the conference proceedings appeared with a delay of about two years.)

\subsection{Discussion on Einstein's report}

The discussion on Einstein's written report was opened by Einstein himself. He begins with the following general statements:

\begin{quote}
We probably all agree that the so-called quantum theory of today is, indeed, a helpful tool but that it is not a theory in the usual sense of the word, at any rate not a theory that could be developed in a coherent form at the present time. On the other hand, it has also turned out that classical mechanics, which finds its expression in the equations of Lagrange and Hamilton, can no longer be viewed as a useful scheme for the theoretical representation of all physical phenomena. ...

This raises the question of which general laws of physics we can still expect to be valid in the domain with which we are concerned. To begin with, we will all agree that the energy principle is to be retained.

In my opinion, another principle whose validity we must retain unconditionally is Boltzmann's definition of entropy through probability. It is to this principle that we owe the faint glimmer of theoretical light we now see shed over the question of states of statistical equilibrium in processes of oscillatory character. But there is still the greatest diversity of opinion as regards the content and domain of validity of this principle. I will therefore first present in brief my view about this matter.
\end{quote}

Einstein goes on with a rather detailed elaboration\footnote{ This has much in common with initial sections of his paper on critical opalescence he had submitted about a year before \cite{Ein7}. For readers, who know German, it is interesting to compare Einstein's presentation with an interesting manuscript for a talk he has given at a meeting of the Physical Society of Z\"{u}rich on November 1910 with the title:  \textit{`Ueber das Boltzmann'sche  Prinzip und und einige aus ihm zu ziehende Folgerungen' }. This document was found only a few years ago (see \cite{pgz}).}, which lead to vivid reactions, in particular by Lorentz, Poincar\'{e}, Wien, Nernst, Langevin and Kamerlingh Onnes. In connection with the fluctuation formula 
(\ref{ein5}), Lorentz emphasized correctly, that the term $h\nu/E=1/Z_q$ is ``totally incompatible with Maxwell's equations and with the prevailing views about electromagnetic processes.'' He supports this conclusion by Einstein with an independent argument.

Generally speaking, Einstein hardly learned anything from the  remarks by his colleagues. I find the comments by Poincar\'{e} particularly disappointing.

\section{Some scattered final remarks}

The scientific program ended with a general debate. This was opened with some brief remarks by Poincar\'{e}. He correctly states that in talks and discussions arguments were often based partly on the old mechanics, but in addition also on hypotheses which are in contradiction to it. This he comments with a statement only a mathematician can make, and which I do not even find funny, namely that it is possible without great trouble to prove any statement, if the proof is based on two contradicting premises. On the bases of such an attitude, quantum mechanics would never have been discovered. The few other comments, in particular by Brillouin, Nernst, Poincar\'{e}, and partly by Langevin, are dominated by rather conservative hopes and statements.

However, this debate was at least much more reasonable in comparison to what Ernest Solvay had to say to the scientists in his final speech. Here, just two examples that illustrate his views on science in general, and of physics in particular:

\begin{quote}
Aber trotzdem, trotz der sch\"{o}nen auf diesem Conseil erzielten Ergebnisse haben Sie die eigentlichen Probleme, die gegenw\"{a}rtig im Vordergrund stehen, nicht gel\"{o}st, Sie haben noch keinen gangbaren Weg er\"{o}ffnet zur exakten Bestimmung der einfachsten Grundelemente, die man vom philosophischen Standpunkt als die eigentlichen Bausteine des aktiven Universums anzusehen hat und auf die ich pers\"{o}nlich ganz besonders meine Untersuchungen gerichtet habe. Auch muss ich Ihnen gestehen, dass meine bisherigen Anschauungen, die ich Ihnen in meiner Er\"{o}ffnungsansprache andeutete, keinerlei Aenderungen erfahren haben. (...)

Inzwischen m\"{o}chte ich noch dem Wunsche Ausdruck verleihen, dass die Versuche verwirklicht werden m\"{o}gen, die auf die Erforschung des Ursprungs der Energie der Brownschen Bewegung und der Energie der Radioaktivit\"{a}t hinzielen. Ich bin fest \"{u}berzeugt, dass diese Energie nicht aus dem Medium stammt, in dem die Brownsche Bewegung vor sich geht und in dem die radioaktiven K\"{o}rper sich befinden, sondern ihren Ursprung ausserhalb desselben hat. (...).\footnote{``But in spite of the beautiful results achieved at this congress, you have not solved the real problems that remain at the forefront. You have not yet opend a practicable path for the exact determination of the simplest basic elements, which one has to regard from a philosophical point of view as the proper building blocks of the active Universe, and to which I personally have particularly directed my investigations. And I must tell you that my prevailing views, which I have indicated to you in my opening speech, have not changed at all.

Meanwhile, I would like to give expression to the desire that attempts will be realized, which call for research towards the search of the origin of the energy of Brownian motion, and the energy of radioactivity. I am firmly convinced that this energy does not originate from the medium in which Brownian motion takes place or in which radioactive bodies are situated, but have their origin outside of it.''}

\end{quote}

Since the German edition of the proceedings appeared with a delay of about two years, the editor -- Arnold Eucken -- added a detailed appendix on the development in the field since the conference, both experimentally and theoretically. By far the most important theoretical contributions were the theories of Debye \cite{debye2}, and -- almost simultaneously -- of Born and Karman \cite{born1}, \cite{born2} on the specific heat of solids. While every student of physics is familiar with Debye's work, the paper of Born and Karman is less known, but more realistic. The two authors did not make the continuum approximation of Debye, but treated the spectrum of lattice vibrations in more detail by making use of work by Madelung. 

Further progress on the specific heat of diatomic molecules has already been indicated. It still took some time until people understood, that at low temperatures the rotationel degrees of freedom are frozen in, a great triumph of the early quantum theory.\footnote{A full understanding of the specific heat of diatomic molecules became only possible with the advent of the new quantum mechanics. Especially for molecular hydrogen, things were only settled after Heisenberg had predicted in 1927 -- on the basis of Pauli's exclusion principle -- two distinct species, called ortho and para hydrogen, that interact only very weakly.}

Before the conference, Einstein dubbed in a letter to Besso \cite{Ein8} the upcoming meeting ``the witch's Sabbath'' and complained ``My twaddle for the Brussels conference weighs down on me.'' After the conference, he wrote to Heinrich Zangger \cite{Ein9} ``Planck stuck stubbornly to some undoubtedly wrong preconceptions'', and he  dismissed Poincar\'{e} with `` Poincar\'{e} was simply negative in general, and, all his acumen notwithstanding, he showed little grasp of the situation.'' In another letter to Besso, Einstein gave low marks to the meeting \cite{Ein10}: ``The congress in Brussels resembled the lamentations on the ruins of Jerusalem'', and ``nothing positive came out of it.'' Einstein was most impressed by H. A. Lorentz, ``ein Wunder von Intelligenz und feinem Takt. Ein lebendiges Kunstwerk. Er ist nach meiner Meinung immer noch der intelligendeste unter den anwesenden Theoretikern gewesen.'' ( Letter to H. Zangger from 15 November, 1911; CPAE, Vol. 5, Doc. 305.)\footnote{[Lorentz] ``is a marvel of intelligence and exquisite tact. A living work of art. In my opinion, he was among the theoreticians present still the most intelligent.''} 

Just as the Solvay Conference was getting under way, the romance between the widowed Marie Curie and Paul Langevin became public. This was, of course, more intersting to the public than anything else, especially because at that very moment it was announced that Madame Curie had won the Nobel Prize in chemistry. After the furor Einstein wrote a gracious letter to her \cite{Ein11}.

I conclude with personal remarks, that look totally disconnected with what was said in this historical account, and may just reflect my advanced age. One often hears that the present day situation in fundamental physics (string theory, loop gravity) has some similarity with the early years of quantum theory, before the great breakthrough -- mostly by a young generation -- in 1925-26. I find this analogy totally wrong. Without the precision experiments by the Berlin group (Kurlbaum, Rubens, etc.) and the difficult measurements of the specific heat of molecular hydrogen and other diatomic gases at low temperatures,  that demonstrated the freezing out of the rotational degrees of freedom, as well as the low temperature measurements of the specific heat of solids by Nernst, Lindemann and others, it is hard to imagine that quantum theory could have been developed. This is, of course, not new, but it may not be inappropriate to be recalled in an article for this journal.

\section*{Acknowledgements}

I am very grateful to Domenico Giulini for detailed constructive criticism of an earlier version of the manuscript, and clarifying discussions.  Thanks go to Thibault Damour for useful suggestions and hints to the literature on the Solvay meetings. I thank G\"{u}nther Rasche for a careful reading of the manuscript.

\end{document}